# A CRYPTOGRAPHIC SCHEME OF MELLIN TRANSFORM

Yeray Cachón Santana

e-mail: ycachon@gmail.com

## ABSTRACT

In this paper it has been developed an algorithm for cryptography, using the Mellin's transform. Cryptography is very important to protect data to ensure that two people, using an insecure channel, may communicate in a secure way. In the present age, ensure the communications will essential to shared data that have to be protected. The original message is a plain text, and the encrypted form as cipher text. The cipher text message contains all the information of the plain text, but is cannot be read from a human without a key and a method to decrypt it.

## 1 INTRODUCTION

The objective of cryptography is enable two people to communicate in a secure way over an unsecure channel. Nowadays, cryptography is essential to protect data and it's used in several internet protocols (SSL...). The cryptography uses an encryption, that allows generating an unintelligible text from a message, and a decryption, that allows recovering the message. So, the goal is that nobody, except the sender and the receiver, could read the message. To ensure that the receiver could recover the message, a shared key must be known only by the sender and the receiver; this shared key will allow encrypting the message by the sender and decrypting it by the receiver.

Changing the key, the cyphertext will be different for the same plaintext so, recover the plaintext without the key is highly improbable.

# 2 MELLIN's TRANSFORM

Let F(x) a function defined for all positive values of t, then the Mellin Transform of f(x) is defined by $f(x)^*(s) = \int_0^\infty f(x) x^{s-1} dx$. This trasnsform has some properties:

1. Scaling:

$$f^*(at)(s) = a^{-s} f^*(s)$$

2. Multiplication by $x^a$

$$(s^a f)^*(s) = \int_0^\infty f(x) x^{(s+a)-1} dx = f * (s+a)$$

3. Inverse of independent variable:

$$(x^{-1} f(x^{-1}))^* = f^*(1-s)$$

4. Multiplication by Power of ln x:

$$\left((\ln x)^k (f(x))\right)^* = \frac{d^k}{ds^k} f * (s)$$

5. Derivate:

$$\left(\frac{d^k}{dx^k} f(x)\right)^* = (-1)^k (s-k)_k f^*(s-k)$$

Where:

$$(s-k)_k = (s-k)(s-j+1)\cdots(s-1) = \frac{\Gamma(s)}{\Gamma(s-k)}$$

6. Convolution

$$(f(x)g(x))^* = \frac{1}{2\pi i} \int_{c-i\infty}^{c+i\infty} F(z) G(s-z) dz$$

# 3 METHOD

Supposing a plain text PT="....", we will consider a polynomial function that contains several constants for each monomial. So, let's consider:

$$L(x) = e^{-x}(G_1 x + G_2 x^2 + \cdots + G_n x^n) = e^{-x} \sum_{i=1}^{N} G_i x^i$$

Where $G_i$ are constants equals to each alphabet's character for the plaintext. For example, with the plaintext "HELLO", each of the constants' values will be:

| chaintext | H | E | L | L | O |
|---|---|---|---|---|---|
| values | 8 | 5 | 12 | 12 | 15 |
| | $G_1$ | $G_2$ | $G_3$ | $G_4$ | $G_5$ |

In this case, the generating function will be:

$$L(x) = e^{-x}(8x + 5x^2 + 12x^3 + 12x^4 + 15x^5)$$

Firstly, let's see how to encrypt using Mellin's transform in general case. We've seen that the generating function is:

$$L(x) = e^{-x}(G_1 x + G_2 x^2 + \cdots + G_n x^n) = e^{-x} \sum_{i=1}^{N} G_i x^i$$

Let's write it in this way:

$$L(x) = e^{-x}(G_1 x + G_2 x^2 + \cdots + G_n x^n) = e^{-x} \sum_{i=1}^{N} G_i x^i = \sum_{i=1}^{N} e^{-x} G_i x^i$$

Using Mellin's transform to each member:

$$L_i^*(x) = \left(e^{-x} G_i x^i\right)^* = \int_0^\infty dx\, e^{-x} x^{s-1} G_n x^n = G_n \int_0^\infty dx\, e^{-x} x^{s-1+n}, \text{ where } s, n \epsilon \mathbb{N}$$

Recalling the Gamma function,

$$\Gamma(x) = \int_0^\infty dx e^{-x} x^{n-1}$$

If $n \in \mathbb{N}$, $\Gamma(n)=(n-1)!$

In this case, we can write the Mellin's transform of each member in this way:

$$L(x)^* = \sum_n L_n^* = \sum_n G_n(s+n-1)! = \sum_n G'_n$$

And, hence, $G'_n = G_n(s+n-1)!$

# ENCRYPTION

So, using the Mellin's transform of a polynomial function, we've deduced the values of each monomial by Mellin's transform. Let's see now what would happen if we choose s=4 for the plaintext "HELLO":

The $G'_i$ will be:

$G'_1 = 8s!$

$G'_2 = 5*(s+1)!$

$G'_3 = 12*(s+2)!$

$G'_4 = 12*(s+3)!$

$G'_5 = 15*(s+4)!$

Choosing s=4, the values will be:

$G'_1 = 192$

$G'_2 = 600$

$G'_3 = 8640$

$G'_4 = 60480$

$G'_5 = 604800$

With those values, the function L(x)*:

$$L(x)^* = e^{-x}(192x + 600x^2 + 8640x^3 + 60480x^4 + 604800x^5)$$

And that will be the encrypted message of the plaintext "HELLO".

Now, let's calculate the message and the private key. For each of the previous values, we will calculate the message as the rest of the division between the number and 26, and the private key will be the quotient of it:

$$192 = 7 * 26 + 10$$

$$600 = 23*26 + 2$$

$$8640 = 332*26 + 8$$

$$60480 = 2326*26 + 4$$

$$604800 = 23261*26 + 14$$

The rests of the divisions are 10,2,8,4,14. If we consider 1=A,2=B,3=C....27=Z, the encrypt message of values {10,2,8,4,14} will be "JBHDN".

So the encrypt message will be "JBHDN" with the private key {4,7,23,332,2326,23261} (It has to be included the value of s on the private key)

**Note**: If this case, because s=4, the values are s,s+1,s+2,s+3,s+4. If the plaintext would be larger than 4, the values would be s,s+1,s+2,s+3,s+4,s,s+1,s+2…

# DECRYPTION

The receiver will share the private key {4,7,23,332,2326,23261} and he will receive "JBHDN". Because he has the private key, he will be able to decrypt the message as a reserve method. Let's see how:

The receiver receives the message "JBHDN" that will correspond to values:

       10 2 8  4 14

       J  B H D N

Firstly, the receiver will recover the $L^*(x)$ function:

| | |
|---|---|
| "J";7 | 7*26+10=192 |
| "B";23 | 23*26+2=600 |
| "H";332 | 332*26+8=8460 |
| "D";2326 | 2326*26+4=60480 |
| "N";23261 | 23261*26+1=604800 |

So, the L*(x) function will be recovered:

$$L^*(x) = e^{-x}(192x + 600x^2 + 8640x^3 + 60480x^4 + 604800x^5)$$

Once that the L*(x) function will be recovered, because the first value of the private key is shared with the sender, we will be able to recover the plain text as the reverse method:

We've seen that the $G'_i$ will be:

$G'_1$=8s!

$G'_2$=5*(s+1)!

$G'_3$=12*(s+2)!

$G'_4$=12*(s+3)!

$G'_5$=15*(s+4)!

In this case, the $G_i$'s have been recovered, and, to decrypt, the plaintext must be found. Let be called $P_i$ the plaintext's values for each value:

$G'_1 = G_1 s!$

$G'_2 = G_2*(s+1)!$

$G'_3 = G_3*(s+2)!$

$G'_4 = G_4*(s+3)!$

$G'_5 = G_5*(s+4)!$

And, hence:

$$G_1 = \frac{G'_1}{s!}$$

$$G_2 = \frac{G'_2}{(s+1)!}$$

$$G_3 = \frac{G'_3}{(s+2)!}$$

$$G_4 = \frac{G'_4}{(s+3)!}$$

$$G_5 = \frac{G'_5}{(s+4)!}$$

As s=4, and the values has been calculated, the plaintext can be recovered:

$$G_1 = \frac{G'_1}{s!} = \frac{192}{4!} = 8$$

$$G_2 = \frac{G'_2}{(s+1)!} = \frac{600}{5!} = 5$$

$$G_3 = \frac{G'_3}{(s+2)!} = \frac{8640}{6!} = 12$$

$$G_4 = \frac{G'_4}{(s+3)!} = \frac{60480}{7!} = 12$$

$$G_5 = \frac{G'_5}{(s+4)!} = \frac{604800}{8!} = 15$$

That corresponds to "HELLO" that is the plaintext ciphered (8 → H; 5 → E; 12 → L; 15 → O)

## ■ EXAMPLE 2

Let's see now for the plaintext "HELLO " if s = 3:

Encryption:

$$G'_1 = 8s!$$

$$G'_2 = 5*(s+1)!$$

$$G'_3 = 12*(s+2)!$$

$$G'_4 = 12*(s+3)!$$

$$G'_5 = 15s!$$

With s=3:

$$G'_1 = 48$$

$$G'_2 = 120$$

$$G'_3 = 1440$$

$$G'_4 = 8640$$

$$G'_5 = 90$$

And, then,

$$L^*(x) = e^{-x}(48x + 120x^2 + 1440x^3 + 8640x^4 + 90x^5)$$

The message and the private key:

48 = 26 + 22

120 = 4*26 + 16

1440 = 55*26 + 10

8640 = 332*26 + 8

90 = 3*26 + 12

So, the message will be "VPJHC" with the private key {3,1,4,55,332,3}

Now, let's decipher the cyphertext "VPJHC" with the private key {3,1,4,55,332,3}. As the previous one, firstly the receiver will recover L*(x):

"V";1                 1*26 + 22 = 48

"P";4                 4*26 + 16 = 120

"J";55                55*26 + 10 = 1440

"H";332               332*26 + 8 = 8640

"C";3                 3*26 + 12 = 90

So,
$$L^*(x) = e^{-x}(48x + 120x^2 + 1440x^3 + 8640x^4 + 90x^5)$$

Because s=3, and the relationship between $G_i$ and $G'_i$ are

$$G_1 = \frac{G'_1}{s!}$$

$$G_2 = \frac{G'_2}{(s+1)!}$$

$$G_3 = \frac{G'_3}{(s+2)!}$$

$$G_4 = \frac{G'_4}{(s+3)!}$$

$$G_5 = \frac{G'_5}{s!}$$

As s=3, and the values has been calculated, the plaintext can be recovered:

$$G_1 = \frac{G'_1}{s!} = \frac{48}{3!} = 8$$

$$G_2 = \frac{G'_2}{(s+1)!} = \frac{120}{4!} = 5$$

$$G_3 = \frac{G'_3}{(s+2)!} = \frac{1440}{5!} = 12$$

$$G_4 = \frac{G'_4}{(s+3)!} = \frac{8640}{6!} = 12$$

$$G_5 = \frac{G'_5}{s!} = \frac{90}{3!} = 15$$

That corresponds to "HELLO" that is the plaintext ciphered (8 → H; 5 → E; 12 → L; 15 → O).

So, with the same plaintext, changing the key s, we'll obtain different cyphertexts but, of course, the decryption of each of them will give the same plaintext for a given key.

## 4 CONCLUSSIONS

In this work a new cryptographic scheme is showed using the Mellin Transform with a private key. For an eyedropper, it will be very difficult to trace the message and decipher it without a private key that allows deciphering the message.